\documentclass[twocolumn,english,aps,pra]{revtex4}
\usepackage[T1]{fontenc}
\usepackage[latin9]{inputenc}
\usepackage{amsmath}
\usepackage{amssymb}
\usepackage{graphicx}
\usepackage{esint}

\makeatletter
\@ifundefined{textcolor}{}
{%
 \definecolor{BLACK}{gray}{0}
 \definecolor{WHITE}{gray}{1}
 \definecolor{RED}{rgb}{1,0,0}
 \definecolor{GREEN}{rgb}{0,1,0}
 \definecolor{BLUE}{rgb}{0,0,1}
 \definecolor{CYAN}{cmyk}{1,0,0,0}
 \definecolor{MAGENTA}{cmyk}{0,1,0,0}
 \definecolor{YELLOW}{cmyk}{0,0,1,0}
}

\usepackage{babel}

\usepackage{babel}

\usepackage{babel}

\usepackage{babel}

\usepackage{babel}

\usepackage{babel}

\makeatother

\usepackage{babel}
\begin{document}
%%%%%%%%%%%%%%%%%%%%%%%%%%%%%% Makra

\global\long\def\r{\left(\mathbf{r}\right)}

\global\long\def\rp{\left(\mathbf{r}'\right)}

\global\long\def\rt{\left(\mathbf{r}\tau\right)}

\global\long\def\rtp{\left(\mathbf{r}\tau'\right)}

\global\long\def\rpt{\left(\mathbf{r}'\tau\right)}

\global\long\def\rptp{\left(\mathbf{r}'\tau'\right)}

\global\long\def\Sh{\hat{\mathbf{S}}}

\global\long\def\rtrptp{\left(\mathbf{r}\tau;\mathbf{r}'\tau'\right)}

\global\long\def\dSh{\cdot\hat{\mathbf{S}}}

\global\long\def\n{\mathbf{n}}

\global\long\def\np{\dot{\mathbf{n}}}

\global\long\def\nnp{\mathbf{n}\cdot\dot{\mathbf{n}}}

\global\long\def\nknp{\mathbf{n}\times\dot{\mathbf{n}}}

\global\long\def\nh{\hat{n}}

\global\long\def\a{\alpha}

\global\long\def\b{\beta}

\global\long\def\s{\sigma}

\global\long\def\D{\Delta}

\global\long\def\sp{\sigma'}

\global\long\def\sb{\sigma"}

\global\long\def\Lb{{\bf \Lambda}}

\global\long\def\wm{\omega_{m}}

\global\long\def\wn{\omega_{n}}

\global\long\def\wl{\omega_{\ell}}

\global\long\def\wmp{\omega_{m'}}

\global\long\def\wnp{\omega_{n'}}

\global\long\def\kwm{\left(\mathbf{k}\omega_{m}\right)}

\global\long\def\uz{U_{0}}

\global\long\def\ud{U_{2}}

\global\long\def\d{\partial}

\global\long\def\e{\epsilon}

\global\long\def\mcS{\mathcal{S}}

\global\long\def\mcD{\mathcal{D}}

\global\long\def\mcH{\mathcal{H}}

\global\long\def\Ek{E_{\mathbf{k}}}

\global\long\def\zz{\left(\mathbf{0}0\right)}

%%%%%%%%%%%%%%%%%%%%%%%%%%%%%%%%%%%%

\title{Finite temperature superfluid transition of strongly-correlated lattice
bosons \\
 in various geometries}

\author{T. A. Zaleski\footnote{Corresponding author. Tel.: +48 713435021; fax: +48 713441029. \textit{E-mail address}: t.zaleski@int.pan.wroc.pl (T.Zaleski).},
T. K. Kope\'{c}}

\affiliation{Institute for Low Temperature and Structure Research Polish Academy
of Sciences,\\
 POB 1410, 50-950 Wroc\l aw 2, Poland}
\begin{abstract}
We study finite-temperature properties of the strongly interacting
bosons in three-dimensional lattices by employing the combined Bogoliubov
method and the quantum rotor approach. Based on the mapping of the
Bose-Hubbard Hamiltonian of strongly interacting bosons onto U(1)
phase action, we study their thermodynamic phase diagrams for several
lattice geometries including; simple cubic, body- as well as face-centered
lattices. The quantitative values for the phase boundaries obtained
here may be used as a reference for emulation of the Bose-Hubbard
model on a variety of optical lattice structures in order to demonstrate
experimental-theoretical consistency for the numerical values regarding
the location of the critical points. 
\end{abstract}


\maketitle

\section{Introduction}

It is well known that the ground state of a system of repulsively
interacting bosons in a periodic potential can be either in a superfluid
state or in a Mott-insulting state, characterized by integer boson
densities and the existence of a gap for particle-hole excitations
\cite{bemodel1}. One key piece of evidence for the Mott insulator
phase transition is the loss of global phase coherence of the matter
wave function. However, there are many possible sources of phase decoherence
in these systems. Substantial decoherence can be induced by quantum
or thermal depletion of the condensate. Experimentally, an enormous
progress was made in the experimental study of cold atoms in optical
lattices \cite{optical_lattices}. Cold atoms interacting with a spatially
modulated optical potential resemble in many respects electrons in
ion-lattice potential of a solid crystals. However, optical lattices
have several advantages with respect to solid state systems. They
can be made to be largely free from defects and can be controlled
very easily by changing the laser field properties. Finally, ultra-cold
atoms confined in optical lattice structure provide a very clean experimental
realization of a strongly correlated many-body problem \cite{quantum_many_body}.
Moreover, in contrast to solids, where the lattice spacings are generally
of order of Angstrom units, the lattice constants in optical lattices
are typically three order of magnitude larger. Furthermore, variety
of multi-dimensional lattices can be experimentally obtained by appropriate
setup of laser beams including cubic face-centered and body-centered
lattices \cite{optlatcryst,optlatbcc}. For example, a three dimensional
(3D) lattice can be created by the interference of at least six orthogonal
sets of counter propagating laser beams. Although the initial system
can be prepared at a relatively low temperature, the ensuing system
after ramp-up of the lattice has a temperature which is usually higher
due to adiabatic and other heating mechanisms. Recent experiments
have reported temperatures on the order of $k_{B}T\sim0.9t$ where
$t$, the hopping parameter, measures the kinetic energy of bosons
\cite{temperature}. At such temperatures, the effects of excited
states become important, motivating investigations of the the \textit{finite
}temperature phase diagrams, showing the interplay between quantum
and thermal fluctuations.

Therefore, the goal of this paper is to provide a study of the combined
effects of a confining lattice potential and finite temperature on
the state diagram of the Bose-Hubbard model in three dimensions in
strongly correlated regime where the standard Bogoliubov treatment
fails to describe the system and a more general framework is required.
Usually, studies of bosons in optical lattices have been conducted
at zero temperature and in two dimensional systems, dealing with Mott
insulator-superfluid transition. In the present work, we explore the
phase transition from the Mott to the superfluid state in a system
of strongly interacting bosons on a cubic lattice with the chemical
potential and temperature as the control parameters. Furthermore,
we employ the quantum rotor method, which uses the module--phase representation
of strongly correlated bosons. This introduces a conjugate to the
density of bosons U(1) quantum phase variable, which acquires dynamic
significance from the boson-boson interaction. The quantum rotor approach
has been verified with other methods \cite{qrcompare}, like quantum
Monte Carlo \cite{qmc} or DMFT \cite{dmft} giving coinciding results.

The plan of the paper is as follows: in Section II, we introduce the
microscopic Bose-Hubbard model relevant for the description of strongly
interacting bosons. Furthermore, in the following Section, we briefly
present technical aspects our quantum rotor approach and in Section
IV we calculate the temperature phase diagrams. Finally, we conclude
in the Section V.

\section{Model Hamiltonian}

The simplest non trivial model that describes interacting bosons in
a periodic potential is the Bose Hubbard Hamiltonian. It includes
the main physics that describe strongly interacting bosons, which
is the competition between kinetic and interaction energy. The realization
of the Bose-Hubbard Hamiltonian using optical lattices has the advantage
that the interaction matrix element $U$ and the tunneling matrix
element $t$ can be controlled by adjusting the intensity of the laser
beams. Its Hamiltonian in a second quantized form reads:\cite{bemodel1}
\begin{eqnarray}
\mathcal{H} & = & -t\sum_{\left\langle \mathbf{r},\mathbf{r}'\right\rangle }\left[a^{\dagger}\left(\mathbf{r}\right)a\left(\mathbf{r}'\right)+a^{\dagger}\left(\mathbf{r}'\right)a\left(\mathbf{r}\right)\right]\nonumber \\
 &  & +\frac{U}{2}\sum_{\mathbf{r}}n^{2}\r-\overline{\mu}\sum_{\mathbf{r}}n\left(\mathbf{r}\right).\label{eq:mainham}
\end{eqnarray}
The first term is the kinetic energy of bosons moving in a given lattice
within a tight-binding scheme, where $t$ represents nearest neighbors
tunneling matrix, $\mathbf{r}$ and $\mathbf{r}'$ are lattice sites
and $\left\langle \mathbf{r},\mathbf{r}'\right\rangle $ denotes summation
over nearest neighbors. The following introduces inter-bosonic correlations
with $U$ being the strength of the on-site repulsive interaction
of bosons. Furthermore, $\overline{\mu}=\mu+\frac{U}{2}$, where $\mu$
is a chemical potential controlling the average number of bosons.
The operators $a^{\dagger}\left(\mathbf{r}\right)$ and $a\left(\mathbf{r}'\right)$
create and annihilate bosons, while the boson number operator $n\r=a^{\dagger}\left(\mathbf{r}\right)a\r$
and a total number of sites is equal to $N$. The Hamiltonian and
its descendants have been widely studied within the last years. The
phase diagram and ground-state properties include the mean-field ansatz,\cite{bemodel1}
strong coupling expansions,\cite{key-2,key-3,key-4} the quantum rotor
approach,\cite{key-5} methods using the density matrix renormalization
group DMRG,\cite{key-6,key-7,key-8,key-9} and quantum Monte Carlo
QMC simulations.\cite{key-10,key-11,key-12,key-13}

\section{U(1) Quantum Rotor Formulation\label{sec:U(1)-Quantum-Rotor}}

The quartic form of the Hamiltonian makes it very difficult to deal
with it in all the different regimes. The aim of this chapter is to
rewrite it so that a systematic approach can be developed to accommodate
strongly interacting regime, In the following, we use a theory that
goes beyond the simple Bogoliubov approximation which has been recently
developed that incorporates the phase degrees of freedom via the quantum
rotor approach to describe regimes beyond the very weakly interacting
one \cite{zaleski_tof}. This scenario provided a picture of quasi-particles
and energy excitations in the strong interaction limit, where the
transition between the superfluid and the Mott state is be driven
by phase fluctuations. Taking advantage of the macroscopically populated
condensate state, we have separated the problem into the amplitude
of the Bose field and the fluctuating phase that was absent in the
original Bogoliubov problem \cite{bogoliubov}.

The statistical sum of the system defined by Eq. \eqref{eq:mainham}
can be written in a path integral form with use of complex fields,
$a\rt$ depending on the ``imaginary time'' $0\le\tau\le\beta\equiv1/k_{B}T$,
(with $T$ being the temperature) that satisfy the periodic condition
$a\rt=a(\mathbf{r}\tau+\beta)$: 
\begin{equation}
Z=\int\left[\mathcal{D}\overline{a}\mathcal{D}a\right]e^{-\mathcal{S}\left[\overline{a},a\right]},\label{statsum}
\end{equation}
where the action $\mcS$ is equal to: 
\begin{equation}
\mathcal{S}\left[\overline{a},a\right]=\int_{0}^{\beta}d\tau\mathcal{H}\left(\tau\right)+\mathcal{S_{B}}\left[\overline{a},a\right],
\end{equation}
where the Berry term is: 
\[
\mathcal{S_{B}}\left[\overline{a},a\right]=\sum_{\mathbf{r}}\int_{0}^{\beta}d\tau\overline{a}\left(\mathbf{r}\tau\right)\frac{\partial}{\partial\tau}a\left(\mathbf{r}\tau\right).
\]
Now, we are briefly introducing the quantum rotor approach.\cite{polak}
The fourth-order term in the Hamiltonian in Eq. (\ref{eq:mainham})
can be decoupled using the Hubbard-Stratonovich transformation with
an auxiliary field $V\rt$: 
\begin{eqnarray}
 &  & e^{-\frac{U}{2}\sum_{\mathbf{r}}\int_{0}^{\beta}d\tau n^{2}\left(\mathbf{r}\tau\right)}\nonumber \\
 &  & \,\,\,\,\,\propto\int\frac{\mathcal{D}V}{\sqrt{2\pi}}e^{\sum_{\mathbf{r}}\int_{0}^{\beta}d\tau\left[-\frac{V^{2}\left(\mathbf{r}\tau\right)}{2U}+iV\left(\mathbf{r}\tau\right)n\left(\mathbf{r}\tau\right)\right]}.
\end{eqnarray}
The fluctuating ``imaginary chemical potential'' $iV\rt$ can be
written as a sum of static $V_{0}\left(\mathbf{r}\right)$ and periodic
function: 
\begin{eqnarray}
V\left(\mathbf{r}\tau\right) & = & V_{0}\left(\mathbf{r}\right)+\delta V\left(\mathbf{r}\tau\right),
\end{eqnarray}
where, using Fourier series: 
\begin{eqnarray}
\delta V\left(\mathbf{r}\tau\right) & = & \frac{1}{\beta}\sum_{\ell=1}^{\infty}\delta V\left(\mathbf{r}\omega_{\ell}\right)\left(e^{i\omega_{\ell}\tau}+e^{-i\omega_{\ell}\tau}\right),
\end{eqnarray}
with the Bose-Matsubara frequencies are $\omega_{\ell}=2\pi\ell/\beta$
and $\ell=0,\pm1,\pm2,\dots$.

\subsection{Phase action}

Introducing the U(1) phase field $\phi\rt$ via the Josephson-type
relation: 
\begin{equation}
\dot{\phi}\left(\mathbf{r}\tau\right)=\delta V\left(\mathbf{r}\tau\right)
\end{equation}
with $\dot{\phi}\rt=\d\phi\rt/\d\tau$ we can now perform a local
gauge transformation to new bosonic variables: 
\begin{equation}
a\rt=b\rt e^{i\phi\rt},\label{eq:gauge_transformation}
\end{equation}
where: 
\begin{equation}
\zeta\rt=e^{i\phi\rt}\label{eq:zeta}
\end{equation}
with $\phi\rt$ being U(1) phase variable. Concerning the amplitude
in Eq. (\ref{eq:gauge_transformation}), the operator splits into
a sum: 
\begin{equation}
b\rt=b_{0}+\delta b\rt.
\end{equation}
Since, the strongly correlated limit is dominated by phase fluctuations,
we neglect a contribution coming from $\delta b\rt$ in subsequent
calculations. After the variable transformations the statistical sum
becomes: 
\begin{equation}
Z=\int\left[\mcD\bar{b}\mcD b\right]\left[\mcD\phi\right]e^{-\mcS\left[\bar{b},b,\phi\right]}
\end{equation}
with the action: 
\begin{align}
 & \mathcal{S}\left[\overline{b},b,\phi\right]=\mathcal{S}_{0}\left[\phi\right]+\mathcal{S_{B}}\left[\overline{b},b\right]\nonumber \\
 & -t\sum_{\left\langle \mathbf{r},\mathbf{r}'\right\rangle }\int_{0}^{\beta}d\tau\left[e^{i\phi(\mathbf{r}'\tau)-i\phi\rt}\overline{b}\left(\mathbf{r}\tau\right)b\left(\mathbf{r}'\tau\right)+h.c.\right]\nonumber \\
 & +\sum_{\mathbf{r}}\int_{0}^{\beta}d\tau\left[U\left|b_{0}\right|^{2}-\overline{\mu}\right]\overline{b}\left(\mathbf{r}\tau\right)b\left(\mathbf{r}\tau\right)\label{eq:eq:phase_bosonic_action_generic}
\end{align}
and 
\begin{equation}
\mathcal{S}_{0}\left[\phi\right]=\sum_{\mathbf{r}}\int_{0}^{\beta}d\tau\left[\frac{\dot{\phi}^{2}\rt}{2U}+i\frac{\overline{\mu}}{U}\dot{\phi}\rt\right].\label{eq:phase_bosonic_action_generic}
\end{equation}
The statistical sum can be integrated over the phase or bosonic variables
with the phase or bosonic action: 
\begin{align}
\mcS\left[\phi\right] & =-\ln\int\left[\mcD\bar{b}\mcD b\right]e^{-\mcS\left[\bar{b},b,\phi\right]},
\end{align}
so that: 
\begin{equation}
Z=\int\left[\mcD\phi\right]e^{-\mcS\left[\phi\right]}.\label{eq:statistical_two}
\end{equation}
In performing the integration in Eq. (\ref{eq:statistical_two}) one
should take phase configurations that satisfy the boundary condition
$\phi\left(\mathbf{r}\beta\right)-\phi\left(\mathbf{r}0\right)=2\pi m\left(\mathbf{r}\right)$
and $m\left(\mathbf{r}\right)=0,\pm1,\pm2,\dots$. The phase-only
action from Eq. (\ref{eq:phase_bosonic_action_generic}) can be written
explicitly: 
\begin{align}
\mcS\left[\phi\right] & =\mathcal{S}_{0}\left[\phi\right]+J\sum_{\left\langle \mathbf{r},\mathbf{r}'\right\rangle }\int_{0}^{\beta}d\tau\cos\left[\phi\rt-\phi\rpt\right],
\end{align}
where $J=t\left|b_{0}\right|^{2}$ represents the stiffness for the
phase field.

\subsection{Phase coherence and order parameter}

The superfluid order parameter is defined by: 
\begin{equation}
\Psi_{B}=\left\langle a\rt\right\rangle =\left\langle b\rt\right\rangle \psi_{B}\equiv\left|b_{0}\right|^{2}\psi_{B},
\end{equation}
where $\left\langle \dots\right\rangle $ denotes the averaging over
effective action depending on pertinent variables. However, a nonzero
value of the amplitude $\left\langle b\rt\right\rangle $ is not sufficient
for superfluidity. Also, the U(1) phase variables must become coherent,
which leads to the phase order parameter: 
\begin{equation}
\psi_{B}=\left\langle e^{i\phi\rt}\right\rangle .
\end{equation}
which is equal to zero in the disordered phase (in particular, the
Mott-insulator for $T=0$). We introduce a unimodular scalar field
$\zeta\left(\mathbf{r}\tau\right)=e^{i\phi\left(\mathbf{r}\tau\right)}$
using the identity: 
\begin{align}
1 & \equiv\int\left[\mathcal{D}^{2}\zeta\right]\prod_{\mathbf{r}}\delta\left[\zeta\left(\mathbf{r}\tau\right)-e^{i\phi\left(\mathbf{r}\tau\right)}\right]\nonumber \\
 & \delta\left[\overline{\zeta}\left(\mathbf{r}\tau\right)-e^{-i\phi\left(\mathbf{r}\tau\right)}\right].
\end{align}
This leads us to the partition function: 
\begin{equation}
Z=\int\left[\mathcal{D}^{2}\zeta\right]\delta\left[\sum_{\mathbf{r}}\left|\zeta\left(\mathbf{r}\tau\right)\right|^{2}-N\right]e^{-\mathcal{S}\left[\zeta,\overline{\zeta}\right]},
\end{equation}
where the unimodularity condition was weakened to be fulfilled on
average and is imposed by a Lagrange multiplier $\lambda$ with the
Laplace transform $\delta\left(x\right)=\int d\lambda e^{\lambda x}$.
The action: 
\begin{equation}
\mathcal{S}\left[\zeta,\overline{\zeta}\right]=\frac{1}{\beta N}\sum_{\mathbf{k}\ell}\overline{\zeta}_{\mathbf{k}}\left(\wl\right)\Gamma_{\lambda_{0}}^{-1}\left(\mathbf{k}\omega_{\ell}\right)\zeta_{\mathbf{k}}\left(\wl\right),
\end{equation}
with the propagator: 
\begin{equation}
\Gamma_{\lambda_{0}}^{-1}\left(\mathbf{k}\omega_{l}\right)=\lambda_{0}-J\left(\mathbf{k}\right)+K^{-1}\left(\omega_{\ell}\right).\label{eq:prop}
\end{equation}
The Fourier transform of the inverse of the phase-phase correlator
$K\left(\tau-\tau'\right)=\left\langle e^{i\phi\left(\mathbf{r}\tau\right)-i\phi\left(\mathbf{r}\tau'\right)}\right\rangle $
depending on a single site only with the average respective to the
phase action only (see, Ref. \cite{zaleski_tof}) reads: 
\begin{equation}
K^{-1}\left(\wl\right)=\frac{U}{4}-U\left[v\left(\frac{\overline{\mu}}{U}\right)+\frac{i\wl}{U}\right]^{2},\label{eq:ppcorelator-1}
\end{equation}
while $J\left(\mathbf{k}\right)=2tb_{0}^{2}\varepsilon_{\mathbf{k}}$,
$\varepsilon_{\mathbf{k}}$ is the dispersion of a given lattice,
$b_{0}$ is the bosonic amplitude obtained from minimalizaition of
the Hamiltonian $\partial\mathcal{H}\left(b_{0}\right)/\d b_{0}=0$:
\begin{equation}
b_{0}^{2}=\frac{zt}{U}+\frac{\overline{\mu}}{U},
\end{equation}
$z$ is a lattice coordination number and, finally, $v(x)=x-[x]-1/2$,
with $[x]$ being the floor function, which gives the greatest integer
less than or equal to $x$ resulting from the periodicity of the phase
variable. In the large-$N$ limit, the value of the Lagrange multiplier
$\lambda$ can be determined from the saddle point method 
\begin{equation}
\left.\frac{\partial\mathcal{S}}{\partial\lambda}\right|_{\lambda=\lambda_{0}}=0\label{eq:saddlepoint}
\end{equation}
with the stationary point value $\lambda_{0}$. Explicitly, from Eq.
(\ref{eq:saddlepoint}) it follows that: 
\begin{equation}
1=\left\langle \overline{\zeta}\left(\mathbf{r}\tau\right)\zeta\left(\mathbf{r}\tau\right)\right\rangle =\frac{1}{\beta N}\sum_{\mathbf{k}\ell}\Gamma_{\lambda_{0}}\left(\mathbf{k}\wl\right).\label{eq:unimod}
\end{equation}
However, in the presence of the condensate, in the ordered phase,
the average unimodularity condition in Eq. (\ref{eq:unimod}) is depleted
by the presence of the order parameter so that: 
\begin{equation}
1-\psi_{B}^{2}=\frac{1}{\beta N}\sum_{\mathbf{k}\ell}\Gamma_{\lambda_{0c}}\left(\mathbf{k}\wl\right),
\end{equation}
where the saddle point value $\lambda_{0c}$ at the critical point
and in the ordered phase, is fixed by the condition: 
\begin{equation}
\Gamma_{\lambda_{0c}}^{-1}\left(\mathbf{k}=0,\wl=0\right)=0,
\end{equation}
which physically means the divergence of the inverse of the uniform
static order parameter susceptibility.

Explicitly, summing over Matsubara frequencies, the Eq. (\ref{eq:unimod})
becomes: 
\begin{equation}
1-\psi_{B}^{2}=\frac{U}{4N}\sum_{\mathbf{k}}\frac{\coth\left(\frac{1}{2}\beta\Xi_{\mathbf{k}}^{-}\right)+\coth\left(\frac{1}{2}\beta\Xi_{\mathbf{k}}^{+}\right)}{\Xi_{\mathbf{k}}},\label{eq:critline}
\end{equation}
where: 
\begin{align}
\Xi_{\mathbf{k}} & =U\sqrt{\frac{\lambda_{0}-\lambda_{0c}}{U}+\frac{2t}{U}b_{0}^{2}\left(\varepsilon_{\mathbf{0}}-\varepsilon_{\mathbf{k}}\right)+v^{2}\left(\frac{\overline{\mu}}{U}\right)},\nonumber \\
\Xi_{\mathbf{k}}^{\pm} & =\Xi_{\mathbf{k}}\pm Uv\left(\frac{\overline{\mu}}{U}\right).
\end{align}
In the next Section we explicitly calculate the outcome of the equation
(\ref{eq:critline}) for several three dimensional lattice geometries.

\section{Results}

In this Section, we first specify the corresponding lattice structure
factors defined as 
\begin{equation}
\varepsilon_{\mathbf{k}}^{{\rm X}}=\sum_{\{{\bf d}\}_{{\rm X}}}\cos\left(d_{x}k_{x}+d_{y}k_{y}+d_{z}k_{z}\right),
\end{equation}
where $\{{\bf d}\}_{{\rm X}}$ denotes a set of vectors connecting
a given site of a lattice $X$ and its nearest neighbors. It should
be noted that in this sense the geometry of the lattice results from
locations of bonds between nearest neighbors for a chosen lattice
site (given by $\left\{ \mathbf{d}\right\} _{X}$) rather than just
simply the location of the lattice sites.

\begin{figure}
\includegraphics[scale=0.4]{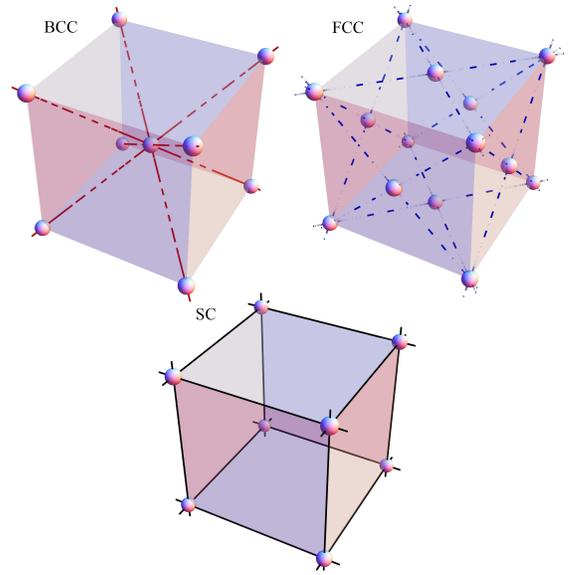}

\protect\caption{\label{fig:Lattices}(Color online) Various geometries of three-dimensional
lattices: SC (simple cubic), BCC (body-centered cubic) and FCC (face-centered
cubic). Bonds between the nearest neighbors are denoted by lines (solid
or dashed) and are located along: the lattice axes (SC), diagonals
of the lattice planes (FCC) and the main diagonals (BCC), respectively.}
\end{figure}

Geometries that we use, are presented in Fig. \ref{fig:Lattices}.
Simple cubic lattice (SC) with the coordination number $z=6$ is defined
by the set of vectors 
\begin{equation}
\{{\bf d}\}_{{\rm SC}}=\{\left(\pm1,0,0\right),\left(0,\pm1,0\right),\left(0,0,\pm1\right)\}
\end{equation}
giving 
\begin{equation}
\varepsilon_{\mathbf{k}}^{{\rm SC}}=\cos k_{x}+\cos k_{y}+\cos k_{z}.
\end{equation}
For face centered lattice (FCC) with the coordination $z=12$ one
has 
\begin{equation}
\{{\bf d}\}_{{\rm FCC}}=\{\left(\pm1,\pm1,0\right),\left(\pm1,0,\pm1\right),\left(0,\pm1,\pm1\right)\}
\end{equation}
and correspondingly 
\begin{equation}
\varepsilon_{\mathbf{k}}^{{\rm FCC}}=2\left(\cos k_{x}\cos k_{z}+\cos k_{x}\cos k_{y}+\cos k_{y}\cos k_{z}\right)
\end{equation}
Furthermore, we consider body centered lattice (BCC), where $z=8$
and $\{{\bf d\}_{{\rm BCC}}}$ is given by 
\begin{equation}
\{{\bf d}\}_{{\rm BCC}}=\{\left(\pm1,\pm1,\pm1\right)\}
\end{equation}
so that 
\begin{equation}
\varepsilon_{\mathbf{k}}^{{\rm BCC}}=4\cos k_{x}\cos k_{y}\cos k_{z}.
\end{equation}

\begin{figure}
\includegraphics[scale=0.65]{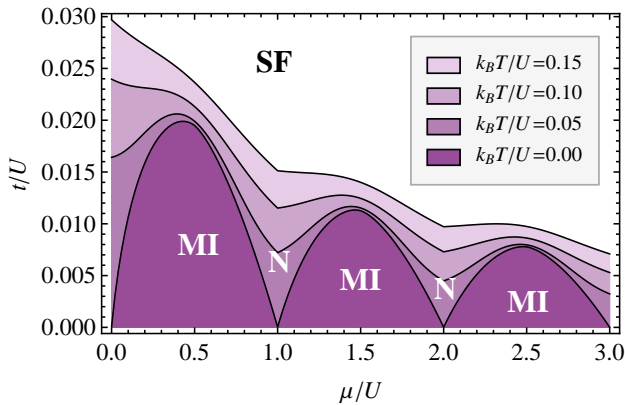}\protect\caption{(Color online) Coupling ratio $t/U$ vs. chemical potential $\mu/U$
phase diagram in various temperatures for the face-centered cubic
lattice (FCC).\label{fig:PhaseDiagFCC}}
\end{figure}

\begin{figure}
\includegraphics[bb=0bp 0bp 360bp 248bp,scale=0.65]{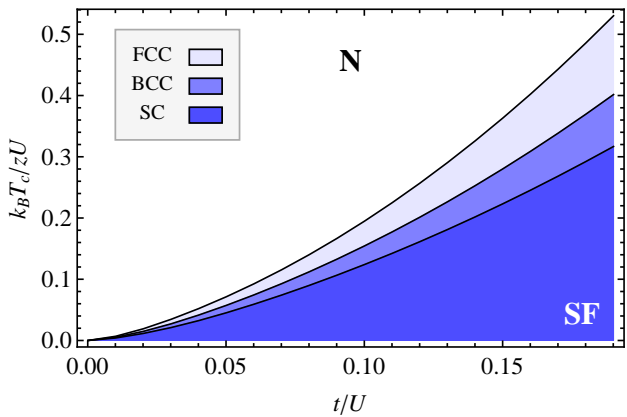}

\protect\caption{(Color online) Critical temperature $k_{B}T/U$ as a function of the
coupling ratio $t/U$ for $\mu/U=1$ and various lattice geometries
(as denoted in the text).\label{fig:T_t}}
\end{figure}

As temperature increases, thermal fluctuations melt away both the
SF and MI phases, introducing the normal (N) phase. For higher temperatures,
the critical coupling $(U/t)_{c}$ for the SF-N transition is lowered
(see, Fig. \ref{fig:PhaseDiagFCC}). With increasing temperature,
the superfluid regions in between the Mott lobes shrink in width and
shift to larger tunneling energies (see, Fig. \ref{fig:PhaseDiagFCC}
for the FCC lattice phase diagram). The phase diagram includes two
different types of phase transition. One type takes place at any generic
point of the phase boundary, and it is driven by the energy cost to
add or subtract small numbers of particles to the incompressible Mott
state as explained above. On the other hand, the other type only occurs
at fixed integer density and takes place at the tip of the lobes.
This transition is driven at fixed density by decreasing $U/t$ and
enabling the bosons to overcome the on site repulsion. The two kinds
of phase transition belong to different universality classes. In the
$T\rightarrow0$ limit, the propagator in Eq. (\ref{eq:prop}) becomes:
\begin{equation}
\Gamma_{\mathbf{k}}^{-1}\left(\wl\right)=r+\mathbf{k}^{2}+\omega_{\ell}^{2}+i\omega_{\ell}+v\left(\frac{\overline{\mu}}{U}\right).\label{eq:propagator}
\end{equation}
Here, $r\sim2tb_{0}^{2}\varepsilon_{\mathbf{0}}-\lambda$ is the critical
``mass'' parameter that vanishes at the phase transition boundary
and $\mathbf{k}^{2}=k\cdot k$. Due to the quantum nature of the problem,
the scaling of the spatial degrees of freedom $\mathbf{k}\rightarrow\mathbf{k}'=s\mathbf{k}$
implies the scaling for frequencies in a form $\wl\rightarrow\omega_{\ell}'=s^{z}\wl$
with the dynamical critical exponent $z$. At the tips of the lobes
in the $t/U$-$\mu/U$ phase diagram (see, Fig. \ref{fig:PhaseDiagFCC}),
one has $v\left(\overline{\mu}/U\right)=0$, so that $\Gamma_{\mathbf{k}}^{-1}\left(\wl\right)\sim k^{2}+\omega_{\ell}^{2}$,
with space-time isotropy giving $z=1$. However, the other points
on the critical line with nonvanishing $v\left(\mu/U\right)$ reflect
the absence of the particle-hole symmetry due to the imaginary term
involving $i\wl$. In this case, the higher order term involving $\omega_{\ell}^{2}$
becomes irrelevant and can be ignored, while the critical form of
the propagator in Eq. (\ref{eq:propagator}) reads $\Gamma_{\mathbf{k}}^{-1}\left(\wl\right)\sim k^{2}+iv\left(\overline{\mu}/U\right)\wl$.
Now, the scaling requires $z=2$ as a result of the momentum-frequency
anisotropy.

\begin{figure}
\includegraphics[scale=0.65]{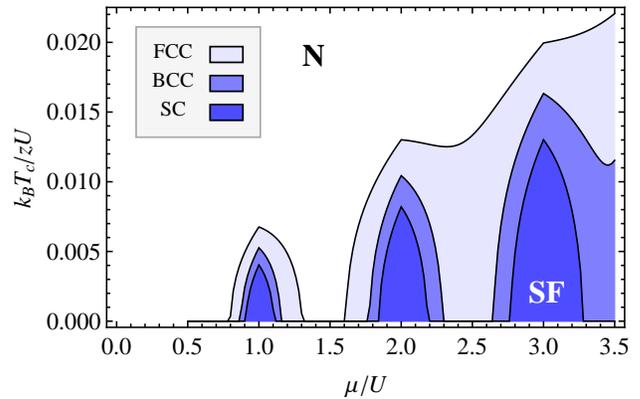}

\protect\caption{(Color online) Dependence of critical temperature $k_{B}T/U$ on the
chemical potential$\mu/U$ for $t/U=0.01$ and various lattice geometries
(as denoted in the text).\label{fig:T_mu}}
\end{figure}

The superfluid critical temperature $T_{c}$ is strongly dependent
on the geometry of the lattice: $T_{c}$ is the highest for the FCC
and is decreasing for BCC and SC lattice, respectively. It can be
also observed in temperature-chemical doping diagrams (see, Fig. \ref{fig:T_mu}):
the FCC lattice requires much higher temperature to destroy the superfluid
phase than the BCC and SC . The ability of the FCC lattice to offer
the highest critical temperature is quite expected. The lattice has
$z=12$ nearest neighbors, as compared to 8 and 6 for BCC and SC lattices,
respectively. We note, that in the mean-field theory, the critical
temperature is simply proportional to the number of the nearest neighbors
\cite{gerbier}. Here, however, the critical temperature is already
normalized by the $z$ factor (see, Figs. \ref{fig:T_t} and \ref{fig:T_mu}).
Therefore our findings, which are based on a more accurate approach,
show that the lattice topology has an additional influence on equilibrium
properties of the Bose-Hubbard model, which are worth to be tested
experimentally.

\section{Conclusions}

In this paper, we have determined the combined effects of three dimensional
lattice potential trapping and temperature for a system of strongly
interacting bosons on several lattice structures. Usually, to be able
to talk about temperature, it is necessary to have a closed system
in thermal equilibrium with a thermal bath. In optical lattices, the
role of the thermal bath is played by the photons. Although, they
do not give the thermal contact and heat exchange required, the reduction
of the entropy of the system is achieved at the expense of the entropy
of the photons. Therefore, the use of the term ``temperature'' is
fully justified. As our calculation exemplify, the finite temperature
equilibrium state is marked by the competing effects of thermally
driven phase fluctuations and phase locking due to hopping of bosons.
Thus, the precise manipulation of this form of matter is of considerable
experimental and theoretical interest. Finally, regarding the theoretical
aspect of our work, it would be also desirable to test of the method
presented by comparing it against numerical solutions obtained by
e.g. diagonalizing the Bose-Hubbard Hamiltonian for a moderate number
of atoms and wells.

\begin{acknowledgments}
We would like to acknowledge support from the Polish National Science
Centre (Grant No. 2011/03/B/ST3/00481).\end{acknowledgments}

\section{Appendix}

Introducing the density of states: 
\begin{equation}
\rho_{{\rm X}}\left(x\right)=\frac{1}{N}\sum_{\mathbf{k}}\delta\left(x-\varepsilon_{\mathbf{k}}^{{\rm X}}\right)
\end{equation}
can greatly simplify numerical calculations, as it converts multiple
sums over wave vectors into a linear integral over a bandwidth {[}e.g.,
in Eq. (\ref{eq:critline}){]}. Here, the index X stands for SC, FCC
or BCC lattices and in several cases a close-form formula for $\rho_{{\rm X}}\left(x\right)$
can be found. Using the dispersion relation from Section IV we enumerate
in the following the relevant cases.

\begin{figure}[h]
\includegraphics[scale=0.67]{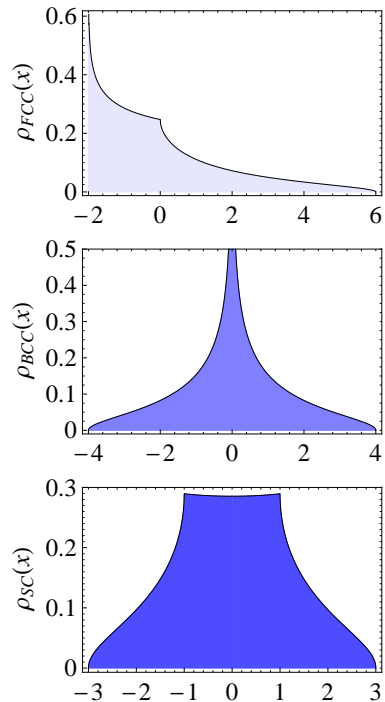}

\protect\caption{\label{denstates}(Color online) Densities of states of the three-dimensional
lattices: simple cubic (SC), face-centered cubic (FCC) and body-centered
cubic (BCC).}
\end{figure}

For the simple cubic lattice one has: 
\begin{eqnarray}
 &  & \rho_{{\rm SC}}\left(x\right)=\frac{1}{\pi^{3}}\int_{\max\left(-1,-2-x\right)}^{\min\left(1,2-x\right)}\frac{du}{\sqrt{1-u^{2}}}\nonumber \\
 &  & \times\mathbf{K}\left[\sqrt{1-\left(\frac{x+u}{2}\right)^{2}}\right]\Theta\left(3-\left|x\right|\right).
\end{eqnarray}
where $\mathbf{K}(x)$ stands for the elliptic integral of the first
kind \cite{abramovitz} and $\Theta(x)$ is the unit step function.
The density of states for the body centered lattice is given by 
\begin{eqnarray}
 &  & \rho_{{\rm BCC}}\left(4x\right)=\frac{1}{2\pi^{3}}\left\{ \mathbf{K}^{2}\left[\sqrt{\frac{1}{2}\left(1+\sqrt{1-x^{2}}\right)}\right]\right.\nonumber \\
 &  & -\left.\mathbf{K}^{2}\left[\sqrt{\frac{1}{2}\left(1-\sqrt{1-x^{2}}\right)}\right]\right\} \Theta\left(1-\left|x\right|\right).
\end{eqnarray}
The formula for the face centered lattice is a bit more involved,
\begin{equation}
\rho_{{\rm FCC}}\left(2x\right)=\frac{1}{2\pi}\lim_{\delta\to0}{\rm Im}G(x-i\delta)
\end{equation}
where $G(x)$ is given by: \\
 for $-1\le x<0$): 
\begin{eqnarray}
 &  & G\left(x\right)=\frac{8}{\pi^{3}\left(x+1\right)}\int_{\arccos\left(\frac{1-x}{2}\right)}^{\arccos\left(\sqrt{-x}\right)}du\mathbf{K}\left[1-k^{2}\left(x,u\right)\right]\nonumber \\
 &  & +\frac{4}{\pi^{3}\left(x+1\right)}\int_{0}^{\arccos\left(\frac{1-x}{2}\right)}\frac{du}{k\left(x,u\right)}\mathbf{K}\left[\frac{\sqrt{k^{2}\left(x,u\right)-1}}{k\left(x,u\right)}\right]\nonumber \\
 &  & +\frac{8}{\pi^{3}\left(x+1\right)}\int_{\arccos\left(\sqrt{-x}\right)}^{\pi/2}du\mathbf{K}\left[\frac{1}{1-k^{2}\left(x,u\right)}\right],
\end{eqnarray}
for $0\le x<1$: 
\begin{eqnarray}
 &  & G\left(x\right)=\frac{8}{\pi^{3}\left(x+1\right)}\int_{\arccos\left(\frac{1-x}{2}\right)}^{\pi/2}du\mathbf{K}\left[1-k^{2}\left(x,u\right)\right]\nonumber \\
 &  & +\frac{4}{\pi^{3}\left(x+1\right)}\int_{0}^{\arccos\left(\frac{1-x}{2}\right)}du\mathbf{K}\left[\frac{1}{1-k^{2}\left(x,u\right)}\right],
\end{eqnarray}
and for $1\le x<3$: 
\begin{eqnarray}
G\left(x\right)=\frac{4}{\pi^{3}\left(x+1\right)}\int_{0}^{\arccos\left(\frac{1-x}{2}\right)}du\frac{\mathbf{K}\left[\frac{\sqrt{k^{2}\left(x,u\right)-1}}{k\left(x,u\right)}\right]}{k\left(x,u\right)},
\end{eqnarray}
where 
\begin{equation}
k\left(x,u\right)=\frac{2\sqrt{x+\cos^{2}\left(u\right)}}{u+1}.
\end{equation}
In Fig. \ref{denstates} we have plotted the outcome for $\rho_{{\rm X}}(x)$
regarding the employed lattices.

\end{document}